\documentclass[12pt,twoside]{article}

\usepackage[latin1]{inputenc}
\usepackage[english]{babel}

\usepackage{latexsym}
\usepackage{amssymb}
\usepackage{amsfonts}
\usepackage{amsmath}

\newtheorem{theorem}{Theorem}
\newtheorem{lemma}{Lemma}
\newtheorem{corollary}{Corollary}
\newtheorem{definition}{Definition}

\newtheorem{proposition}{Proposition}

\date{}

\def\be{\begin{equation}}
\def\ee{\end{equation}}
\def\bc{\begin{center}}
\def\ec{\end{center}}

\begin{document}

\centerline{}

\centerline{}

\centerline {\Large{\bf Dilution Robustness for Mean Field Ferromagnets}}

\centerline{}

\centerline{}

\centerline{\bf {Adriano Barra$^{*\dag}$, Federico Camboni$^{*}$,
Pierluigi Contucci$^{\dag}$}}

\centerline{}

\centerline{$^*$ Dipartimento di Fisica,
    Sapienza Universit\`a di Roma}
\centerline{$^\dag$ Dipartimento di Matematica, Universit\`a di
Bologna}

\centerline{}

\newtheorem{Theorem}{\quad Theorem}[section]

\newtheorem{Definition}[Theorem]{\quad Definition}

\newtheorem{Corollary}[Theorem]{\quad Corollary}

\newtheorem{Lemma}[Theorem]{\quad Lemma}

\newtheorem{Example}[Theorem]{\quad Example}

\begin{abstract}
In this work we compare two different random dilutions on a mean
field ferromagnet: the first model is built on a Bernoulli-diluted
graph while the second lives on a Poisson-diluted one. While it is
known that the two models have, in the thermodynamic limit, the
same free energy, we investigate on the structural constraints
that the two models must fulfill. We rigorously derive for each
model the set of identities for the multi-overlaps distribution,
using different methods for the two dilutions: constraints in the
former model are obtained by studying the consequences of the
self-averaging of the internal energy density, while in the latter
are obtained by a stochastic-stability technique. Finally we prove
that the identities emerging in the two models are the same,
showing {\itshape robustness} of the ferromagnetic properties of
diluted graphs with respect to the details of dilution.
\end{abstract}

{\bf Keywords:} diluted graphs, spin glasses, polynomial
identities.

\section{Introduction}

In the past decade an increasing interest has been shown in
statistical mechanics built on diluted graphs (see i.e.
\cite{barabasi}\cite{barrat}\cite{bucca}\cite{barabba3}\cite{ton3}\cite{smallworld}).
For diluted spin glasses \cite{GTVB}\cite{BarDes} this interest
is, at least, double motivated: despite their mean field nature,
they share with finite-dimensional models the fact that each spin
interact with a finite number of other spins. Moreover they are
mathematically equivalent to some random optimization problems
(i.e. K-SAT or X-OR-SAT depending on the size of the instantaneous
interaction \cite{Ksat1}\cite{Ksat2}).

Although the cases of simpler ferromagnetic models \cite{barra0}\cite{barraP}\cite{ellis}
are not interesting from the point of view of the hard satisfiability interpretation,
they are still interesting for their finite connectivity nature and for testing
different variations on the topology of the graph they are based on.

With this aim we consider two different ways of diluting the graph
\cite{fan}: the first ferromagnet has its links distributed
according to a Bernoulli probability distribution, the
second according to a Poisson.

For these models we compared the properties of a family of linear
constraints for the order parameters (known as
Aizenman-Contucci polynomials \cite{aizcon}\cite{barra1} in the case
of spin glasses).

These relations were investigated earlier \cite{aizcon}\cite{guerra2} in the spin glass
framework, where they were obtained as a result of the stability of the quenched measure with
 respect to random perturbation, or equivalently through the bound on the
fluctuation  of the internal energy. Here we propose them as
 a test for robustness under dilution.

The methods to approach the identities for the two models are
structurally different. For the Poisson case, in fact, the
additive law of Poissonian random variables makes possible the
direct exploitation of the stochastic stability property. The same
strategy is not available for the Bernoulli random variables but,
for those, we derive the set of identities from the general bound
on the quenched fluctuations (even though, for the sake of
completeness, in the Appendix we derive the constraints within
this general framework for the Poisson case too). The methods we
use are generalizations of those appearing in
\cite{barraguerra}\cite{BCK}\cite{CL}\cite{CDGG}\cite{limterm}\cite{pastur}\cite{CGN}\cite{gg}\cite{barra1}\cite{ABC}.

Our main result is a rigorous proof of the identities and, especially,
the fact that they coincide for the two dilutions.

\section{The mean field diluted ferromagnet}

We introduce a large number $N$ of sites, labeled by the index
$i=1,...,N$, and associate to each of them an Ising variable  $\sigma_i=\pm1$.
\newline
We introduce furthermore two families of discrete independent random
variables $\{i_{\nu}\}$,$\{j_{\nu}\}$, uniformly distributed on
 $1,2,...,N$.
 \newline
 The Hamiltonian $H_N(\sigma)$ of the diluted ferromagnet is
 expressed trough
\begin{equation}
H_{N}(\sigma) = - \sum_{\nu=1}^{x} \sigma_{i_\nu}\sigma_{j_\nu}
\end{equation}
where $x$ does depend on the dilution probability distribution.
\newline
For the Bernoullian dilution case the variable  $x$ is called $k$,
and defined by
\begin{equation}
E_B[\cdot] = \sum_{k=0}^{M}\frac{M!}{(M-k)!k!}
(\frac{\alpha}{N})^k(1-\frac{\alpha}{N})^{M-k}[\cdot],
\end{equation}
$M=N(N-1)/2$ being the maximum amount of couples
 $\sigma_i\sigma_j$ existing in the model and $\alpha/N$ the
 probability that two spins interact.
\newline
$\alpha >0$ plays the role of the connectivity.
\newline
The mean and the variance of $k$ are obtained as
\begin{eqnarray}
E_B[k] &=&
\frac{M\alpha}{N}, \\
E_B[k^2] - E_B^2[k] &=& \frac{M\alpha}{N}(1 - \frac{\alpha}{N}).
\end{eqnarray}
\medskip
We will make use in the sequel of following properties of
a Bernoulli distribution:
\begin{eqnarray}
E_B[kg(k)] &=& \frac{M\alpha}{N}E_B[g(k+1)], \label{B1}\\
E_B[k^2g(k)] &=&  \frac{M(M-1)\alpha^2}{N^2}E_B[g(k+2)] -
\frac{M\alpha}{N}E_B[g(k+1)],\label{B2} \\
\frac{d}{d\alpha}E_B[g(k)] &=& \frac{M}{N}E_B[g(k+1) -
g(k)].\label{B3}
\end{eqnarray}

\bigskip

For the Poissonian dilution case, we denote $x$ as $\xi_{\alpha
N}$, which is a random variable of mean $\alpha N$, for
some $\alpha > 0$ (again defining the connectivity of the model),
such that \be P(\xi_{\alpha N} = k) = \pi(k,\alpha N)= \exp\Big(
-\alpha N \Big) \frac{(\alpha N)^k}{k!}, \ k = 0,1,2,...\ee
\newline
Furthermore, we stress that the Poisson distribution obeys the
following properties
\begin{eqnarray}
k \pi(k,\lambda) &=& \lambda \pi(k-1,\lambda), \\
\frac{d}{d\lambda}\pi(k,\lambda) &=&
-\pi(k,\lambda)+\pi(k-1,\lambda)(1-\delta_{k,0}).
\end{eqnarray}
As for the Bernoulli case, the average with respect to the Poisson
measure will be denoted by the index
\begin{equation} E_P =
\sum_{k=0}^{\infty} \frac{e^{\alpha N} (\alpha N)^k}{k!}.
\end{equation}
\newline
The expectation with respect to all the quenched
variables will be denoted by \textbf{E} and represents the product
of the expectation over the dilution distribution and the expectation over the uniformly
distributed variables
$$
\textbf{E} = E_{B,P} \cdot \frac{1}{N^2}\sum_{i,j}^{1,N}.
$$
\newline
The thermodynamic objects we deal with are the partition
function
\begin{equation}
Z_{N}(\alpha,\beta) = \sum_{\{\sigma\}} e^{-\beta H_N(\alpha)},
\end{equation}
the quenched intensive free energy
\begin{equation}
A_{N}(\alpha,\beta) = \frac{1}{N} \textbf{E}\ln
Z_{N}(\alpha,\beta),
\end{equation}
the Boltzmann state
\begin{equation}
\omega(g(\sigma)) = \frac{1}{Z_{N}(\alpha,\beta)}
\sum_{\{\sigma_N\}} g(\sigma) e^{-\beta H_{N}(\alpha)},
\end{equation}
the replicated Boltzmann state
\begin{equation}
\Omega(g(\sigma)) = \prod_s \omega^{(s)}(g(\sigma^{(s)}))
\end{equation}
and the global average $\langle g(\sigma)\rangle$ defined as
\begin{equation}
\langle g(\sigma)\rangle = \textbf{E}[\Omega(g(\sigma))].
\end{equation}
\medskip
\newline
The functional order parameter of the theory is the infinite family
of multi-overlaps, defined as
$$
q_{1\cdots
n}=\frac1N\sum_{i=1}^{N}\sigma^{(1)}_{i}\cdots\sigma^{(n)}_{i} ,
$$
where particular emphasis is  due to  the magnetization $m = q_1 =
(1/N)\sum_{i=1}^{N}\sigma_{i}$ and to the two replica overlap
$q_{12} = (1/N)\sum_{i=1}^{N}\sigma_{i}^1\sigma_i^2$.

\section{Bernoullian diluted case}

Identities in the Bernoullian model will be obtained as a
consequence of the internal energy self-average
\cite{contgia}\cite{gg}; before focusing on this procedure, let us
recall that
\begin{definition}
A quantity $A(\sigma)$ is called {\itshape self-averaging} if \be
\lim_{N \to \infty}\langle \Big(A(\sigma) - \langle
A(\sigma)\rangle\Big)^2 \rangle = \lim_{N \to \infty} \Big(
\langle A^2(\sigma) \rangle - \langle A(\sigma) \rangle^2 \Big) =
 0, \ee
\end{definition}
by which we can remind the
\begin{proposition}
Given two regular functions $A(\sigma)$ and $B(\sigma)$, if at
least one of them is self-averaging, then the following relation
holds \be\label{idea} \lim_{N\to\infty}\langle A(\sigma)B(\sigma)
\rangle = \lim_{N\to\infty} \langle A(\sigma) \rangle
\lim_{N\to\infty}\langle B(\sigma) \rangle\ee
\end{proposition}
\textbf{Proof}
\newline
The proof is straightforward. Let us suppose the self-averaging
quantity is $B(\sigma)$ and use $A(\sigma)$ as a trial function.
Then we have
\begin{eqnarray}
0 &\leq& |\langle A(\sigma)B(\sigma)\rangle-\langle A(\sigma)
\rangle \langle B(\sigma) \rangle| \\ \nonumber &=& |\langle
A(\sigma)B(\sigma) - A(\sigma)\langle B(\sigma) \rangle + \langle
A(\sigma) \rangle B(\sigma)\rangle - \langle A(\sigma) \rangle
\langle B(\sigma) \rangle|
\\ &\leq& \sqrt{\langle
A^2(\sigma)\rangle}\sqrt{\langle (B(\sigma) - \langle B(\sigma)
\rangle)^2 \rangle}, \nonumber
\end{eqnarray}
where, in the last passage, we used Cauchy-Schwartz relation.
\newline
In the thermodynamic the proof becomes completed. $\Box$
\newline
\newline
The scheme to follow is then clear: using the above proposition as
the underlying backbone in the derivation of the constraints in
this section, we must, at first, show that the internal energy
density of the model self-averages. Then we use as trial functions
suitably chosen quantities of the order parameters.
\newline
The identities follow by evaluating explicitly both the terms of
eq. (\ref{idea}): this operation produces several contributions,
all involving the order parameters, among which cancelations occur
and the remaining part gives the identities.

\subsection{Self-averaging of the internal energy density}

Once defined $h_l = H(\sigma^{(l)})/N$ as the density of the
Hamiltonian evaluated on the generic $l^{th}$ replica and
\begin{eqnarray}
\theta &=& \tanh(\beta) \\
\alpha'&=& M\alpha / N^2
\quad \stackrel{N\rightarrow\infty}{\longrightarrow}
\quad \alpha / 2
\end{eqnarray}
for simplicity, we start with the following
\begin{theorem}
In the thermodynamic limit, and in $\beta$-average, the internal
energy density self-averages \be \lim_{N \to \infty}
\int_{\beta_1}^{\beta_2}\mathbb{E}\Big( \Omega(h^2) - \Omega(h)^2
\Big) d\beta = 0. \ee
\end{theorem}
This Theorem has already been proved in \cite{dembo} by using
essentially the existence of the thermodynamic limit for the free
energy density. To make the paper self contained we provide an
alternative proof in the appendix.

\bigskip

We can now introduce the following lemma.
\begin{lemma}
Let us consider for simplicity the quantity
\begin{equation}\label{delta1G}
\Delta G = \sum_{l=1}^{s} \Big[E \big(\Omega(h_lG) -
\Omega(h_l)\Omega(G)\big) \Big].
\end{equation}
For every regular, smooth function $G$, in $\beta$-average, we
have
\begin{equation}
\lim_{N\rightarrow\infty} \int_{\beta_1}^{\beta_2}|\Delta G|d\beta
= 0.
\end{equation}
\end{lemma}
\textbf{Proof}
\begin{eqnarray}
\int_{\beta_1}^{\beta_2}|\Delta G|d\beta &\leq&
\int_{\beta_1}^{\beta_2}\sum_{l=1}^{s}|
\:\textbf{E}[\Omega(h_lG) - \Omega(h_l)\Omega(G)] \:|\:d\beta \label{G1}\\
&\leq& \int_{\beta_1}^{\beta_2}\sum_{l=1}^{s}
\sqrt{\textbf{E}[(\Omega(h_lG) - \Omega(h_l)\Omega(G))^2] }d\beta \label{G2} \\
&\leq& s \int_{\beta_1}^{\beta_2}
\sqrt{\textbf{E}[\Omega(h^2) - \Omega^2(h)] }d\beta \label{G3} \\
&\leq& s \sqrt{\beta_2 - \beta_1}\sqrt{\int_{\beta_1}^{\beta_2}
\textbf{E}[\Omega(h^2) - \Omega^2(h)] d\beta} \;
\stackrel{N\rightarrow\infty}{\longrightarrow}0 \label{G4}
\end{eqnarray}
where (\ref{G1}) comes from triangular inequality; (\ref{G2}) is
obtained via the Jensen inequality applied to the measure
$\textbf{E}[\cdot]$. In the same way (\ref{G3}) comes from Schwarz
inequality applied on the measure $\Omega(\cdot)$ (being $G$
bounded), while (\ref{G4}) is obtained via Jensen inequality
applied on the measure
$(\beta_2-\beta_1)^{-1}\int_{\beta_1}^{\beta_2}(\cdot)d\beta$. \
$\Box$
\newline
\newline
Now we can state the main theorem for the linear constraints.
\newline
We are going to introduce a specific trial functions that
we call $f_G(\alpha,\beta)$.
\newline
\newline
\begin{theorem}\label{T1}
Let us consider the following series of multi-overlap
functions $G$ acting, in complete generality, on $s$ replicas
\begin{eqnarray}\label{F1}
f_G(\alpha,\beta) &=& \alpha' \Big[ \Big(\sum_{l=1}^{s}\langle G
m_l^2 \rangle - s\langle G m_{s+1}^2 \rangle\Big)
\Big(1 - \theta^2\Big) + \nonumber \\
&& + 2\theta \Big(\sum_{a<l}^{1,s}\langle G q_{al}^2 \rangle -
s\sum_{l}^{1,s}\langle G q_{l,s+1}^2 \rangle +
\frac{s(s+1)}{2}\langle G q_{s+1,s+2}^2 \rangle \Big) + \nonumber \\
&& + 3\theta^2\Big(\sum_{l<a<b}^{1,s}\langle G q_{l,a,b}^2 \rangle
- s\sum_{l<a}^{1,s}\langle G q_{l,a,s+1}^2 \rangle +
\frac{s(s+1)}{2}\sum_{l}^{1,s}\langle G q_{l,s+1,s+2}^2 \rangle + \nonumber \\
&& \quad - \frac{s(s+1)(s+2)}{3!}\langle G q_{s+1,s+2,s+3}^2
\rangle  \Big)
 + O(\theta^3)\Big],
\end{eqnarray}
in the thermodynamic limit the following generator of linear
constraints holds:
\begin{equation}\label{f1}
\lim_{N\rightarrow\infty}\int_{\beta_1}^{\beta_2} d\beta
|f_G(\alpha,\beta)| = 0.
\end{equation}
\end{theorem}
\medskip
\textbf{Proof}
\newline
Let us consider explicitly the quantities encoded in
(\ref{delta1G}). For the sake of clearness all the calculations
are reported in appendix, here we present just the results.
\begin{eqnarray}
\textbf{E}[\Omega(h_lG)] &=& -\alpha' \Big[\langle G m_l^2 \rangle
+ \theta\Big(\sum_{a=1}^{s}\langle G q_{a,l}^2 \rangle -
s\langle G q_{l,s+1}^2 \rangle\Big) +  \nonumber \\
&& + \theta^2\Big(\sum_{a<b}^{1,s}\langle Gq_{l,a,b}^2 \rangle -
s\sum_{a}^{1,s}\langle Gq_{l,a,s+1}^2 \rangle +
\frac{s(s+1)}{2}\langle Gq_{l,s+1,s+2}^2 \rangle\Big) \nonumber \\
&& + O(\theta^2)\Big],
\end{eqnarray}
\begin{eqnarray}
\textbf{E}[\Omega(h_l)\Omega(G)] &=& -\alpha' \Big[\langle G
m_{l}^2 \rangle + \theta\Big(\sum_{a=1}^{s+1}\langle G q_{a,l}^2
\rangle -
(s+1)\langle G q_{l,s+1}^2 \rangle\Big) + \nonumber \\
&& + \theta^2\Big(\sum_{a}^{1,s}\langle Gm_{a}^2 \rangle -
(s+1)\langle Gm_{l}^2 \rangle +
\sum_{a<b}^{1,s}\langle Gq_{l,a,b}^2 \rangle + \nonumber \\
&& - (s+1)\sum_{a}^{1,s}\langle Gq_{l,a,s+1}^2 \rangle +
\frac{(s+1)(s+2)}{2}\langle Gq_{l,s+1,s+2}^2 \rangle\Big) \nonumber \\
&& + O(\theta^2)\Big].
\end{eqnarray}
Subtracting the last equation from the former, immediately we
conclude that
\begin{equation}\label{delta1Gesp}
\Delta G = - f_G(\alpha,\beta),
\end{equation}
from which theorem thesis follows. $\Box$

\subsection{Linear constraints for multi-overlaps}

We outline here the first order identities as it is customary to do
in the spin-glasses counterpart
\cite{barra1} or in neural network \cite{BG2}.
\newline
\begin{proposition}
The first class of multi-overlap constraints is obtained by
choosing $G=m^2$.
\end{proposition}
In fact, if we set  $G = q_1^2 = m^2$, the function
$f_G(\alpha,\beta)$ becomes
\begin{eqnarray}\nonumber
f_{m^2}(\alpha,\beta) &=& \alpha'\Big[\Big(\langle m_1^4\rangle -
\langle m_1^2m_2^2\rangle\Big)\Big(1 - \theta^2\Big) + \\
\nonumber && - 2\theta\Big(\langle m_1^2q_{12}^2\rangle - \langle
m_1^2q_{23}^2\rangle\Big) + \\ \nonumber && + 3\theta^2
\Big(\langle m_1^2q_{123}^2\rangle - \langle m_1^2q_{234}^2\rangle
\Big) + O(\theta^3)\Big],
\end{eqnarray}
from which, changing the Jacobian $d\theta = (1 -
\theta^2)d\beta$, we get
\begin{eqnarray}\label{aAC1}
\lim_{N\rightarrow\infty}\int_{\beta_1}^{\beta_2}
|f_{m^2}(\alpha,\beta)|d\beta &=&
\frac{\alpha}{2}\int_{\theta_1}^{\theta_2}d\frac{\theta}{(1 -
\theta^2)} \Big[|\; \Big(\langle m_1^4\rangle - \langle m_1^2m_2^2\rangle\Big) + \\
\nonumber  && - 2\theta \Big(\langle
m_1^2q_{12}^2\rangle - \langle m_1^2q_{23}^2\rangle\Big) + \\
\nonumber && + 3\theta^2 \Big(\langle
m_1^2q_{123}^2\rangle - \langle m_1^2q_{234}^2\rangle \\
\nonumber && + O(\theta^3)\;|\Big] = 0,
\end{eqnarray}
where, the (not interesting) breakdown at $\theta=1$, of the
expression above, reflects the lack of convergence of the
harmonic series we used in eq. (\ref{breakdown}).
\newline
\begin{proposition}
The second class of multi-overlap constraints is obtained by
choosing $G=q_{12}^2$.
\end{proposition}
In fact, if we set $G = q_{12}^2$, the function
$f_G(\alpha,\beta)$ becomes
\begin{eqnarray} \nonumber
f_{q_{12}^2}(\alpha,\beta) &=& \alpha'\Big[ \Big( 2\langle
m_1^2q_{12}^2\rangle - 2\langle m_3^2q_{12}^2\rangle\Big)\Big(1 -
\theta^2\Big) + \\ \nonumber && + 2\theta\Big( \langle
q_{12}^4\rangle - 4\langle q_{12}^2q_{23}^2\rangle + 3\langle
q_{12}^2q_{34}^2\rangle\Big) + \\ \nonumber && - 6\theta^2\Big(
\langle q_{12}^2q_{123}^2\rangle - 3\langle
q_{12}^2q_{234}^2\rangle + 2\langle q_{12}^2q_{345}^2\rangle\Big)
+ O(\theta^3)\Big].
\end{eqnarray}
Again
\begin{eqnarray}\label{aAC2}
\lim_{N\rightarrow\infty}\int_{\beta_1}^{\beta_2}
|f_{q^2}(\alpha,\beta)|d\beta &=&
\frac{\alpha}{2}\int_{\theta_1}^{\theta_2}d\frac{\theta}{(1 -
\theta^2)} \Big[|\; \Big(\langle
m_1^2q_{12}^2\rangle - \langle m_3^2q_{12}^2\rangle\Big) + \\
\nonumber && + \theta \Big( \langle q_{12}^4\rangle - 4\langle
q_{12}^2q_{23}^2\rangle + 3\langle q_{12}^2q_{34}^2\rangle\Big) +
\\ \nonumber && - 3\theta^2 \Big( \langle
q_{12}^2q_{123}^2\rangle - 3\langle q_{12}^2q_{234}^2\rangle +
2\langle q_{12}^2q_{345}^2\rangle\Big) \nonumber \\ \nonumber && +
O(\theta^3)\;|\Big] = 0,
\end{eqnarray}
from which the constraints are obtained as the r.h.s. of
(\ref{aAC1},\ref{aAC2}) set to zero. $\Box$

\section{Poissonian diluted case}

To tackle the Poisson diluted ferromagnet, we are going to use the
cavity field approach.
\newline
The idea beyond this technique is that, calling $F(\beta)$ the
extensive free energy and $f(\beta)$ the intensive one, a bridge
among the two, in the large $N$ limit, is offered simply by the
relation \be \Big(- F_{N+1}(\beta)- F_N(\beta)\Big)= f(\beta) +
O(N^{-1}). \ee As our system has a topologically quenched
disordered the $N+1$ spin can be seen as an ''external random
cavity field'' for the former system of $N$ particles.
\newline
The identities derived by tuning that field are called of
''stochastic stability''. The simplest way to find them is to
consider monomials which are left invariant by the random field:
the derivative with respect to it, being zero, will produce the
desired polynomial.

\subsection{Cavity field decompositions for the pressure density}

To start applying the sketched plan let us decompose (in
distribution) a Poissonian random Hamiltonian of $N+1$ spins in
two Hamiltonians \cite{ABC}: The former of the "inner" $N$
interacting spins, the latter as the remaining spin interacting with
the internal $N$ spins of the cavity.

Up to negligible corrections that go to zero in the thermodynamic limit,
we can write in distribution
\begin{equation}
H_{N+1}(\alpha) = -\sum_{\nu=1}^{P_{\alpha (N+1)}}
\sigma_{i_\nu}\sigma_{j_\nu} \quad \sim \quad
-\sum_{\nu=1}^{P_{\tilde{\alpha} N}} \sigma_{i_\nu}\sigma_{j_\nu}
- \sum_{\nu=1}^{P_{2\tilde{\alpha }}} \sigma_{i_\nu}\sigma_{N+1},
\end{equation}
or simply for compactness
\begin{equation}\label{hsplit}
 H_{N+1}(\alpha) \sim H_{N}(\tilde{\alpha}) +
\hat{H}_{N}(\tilde{\alpha})\sigma_{N+1}
\end{equation}
where
\begin{equation}\label{decompo}
\tilde{\alpha} = \frac{N}{N+1}\alpha \stackrel{N\rightarrow
\infty}{\longrightarrow} \alpha, \qquad
\hat{H}_{N}(\tilde{\alpha}) = - \sum_{\nu=1}^{P_{2\tilde{\alpha}}}
\sigma_{i_\nu}.
\end{equation}
\medskip
\newline
It is useful now to introduce an interpolating parameter $t \in
[0,1]$ on the term encoding the linear connectivity shift so to
menage the derivative with respect to the random field by
differentiating with respect to this parameter.
\begin{definition}
We define the $t$-dependent Boltzmann state $\tilde{\omega}_t$ as
\begin{equation}\label{dante}
\tilde{\omega}_t(g(\sigma)) = \frac{1}{Z_{N,t}(\alpha,\beta)}
\sum_{\{\sigma\}}g(\sigma) e^{\beta\sum_{\nu=1}^{P_{\tilde{\alpha}
N}} \sigma_{i_\nu}\sigma_{j_\nu} + \beta
\sum_{\nu=1}^{P_{2\tilde{\alpha }t}} \sigma_{i_\nu}}.
\end{equation}
\end{definition}
We stress the simplicity by which the $t$ parameter switches among
the system of $N+1$ spins and the one built just by the former
$N$,  in the large $N$ limit: In fact, being the two body
Hamiltonian left invariant by the gauge symmetry $\sigma_i \to
\epsilon\sigma_i$ for all $i\in(1,...,N)$ with $\epsilon = \pm 1$,
by choosing $\epsilon = \sigma_{N+1}$ we have
\begin{eqnarray}
Z_{N,t=1}(\tilde{\alpha},\beta) &=& Z_{N+1}(\alpha,\beta), \\
Z_{N,t=0}(\tilde{\alpha},\beta) &=& Z_{N}(\tilde{\alpha},\beta).
\end{eqnarray}
Note that $Z_{N,t}(\tilde{\alpha},\beta)$ is defined accordingly
to (\ref{dante}) and coherently, dealing with the perturbed
Boltzmann measure, we introduce an index $t$ also to the global
averages $\langle . \rangle \to \langle . \rangle_t$.

\subsection{Stochastic stability via cavity fields}

We are now ready to attack the problem.
\newline
We divide the ensemble of overlap monomials in two large categories:
stochastically stable monomials and (as a side results) not
stochastically stable ones. Then we find explicitly the family of
the stochastically stable monomials, and by putting their
$t$-derivative equal to zero we obtain the identities. To follow
the plan let us introduce the
\begin{definition}
We define as stochastically stable monomials those multi-overlap
monomials where each replica appears an even number of times.
\end{definition}
We are ready to introduce the main theorem, which offers, as a
straightforward consequence, a useful corollary, stated
immediately after.
\newline
\begin{theorem}\label{Tcwdiluito1}
\textit{At $t=1$ the Boltzmannfaktor of the perturbed measure is
comparable with the canonical Boltzmannfaktor, and, in the
thermodynamic limit, we get}
\begin{equation}
\lim_{N \to \infty}
\mathbb{E}\tilde{\Omega}_{N,t=1}(\sigma_{i_1}\sigma_{i_2}...\sigma_{i_n})
=
\lim_{N\to\infty}\mathbb{E}\Omega_{N+1}(\sigma_{i_1}\sigma_{i_2}...\sigma_{i_n}\sigma_{N+1}^n).
\end{equation}
\end{theorem}

\begin{corollary}\label{Tcwdiluito3}
\textit{In the thermodynamic limit, the averages
$\langle\cdot\rangle_t$ of the stochastically stable monomials
become  t-independent in $\beta$-average.}
\end{corollary}
\textbf{Proof}
\newline
Let us focus on the proof of Theorem \ref{Tcwdiluito1}. Corollary
\ref{Tcwdiluito3} will be produced as a straightforward
application of Theorem \ref{Tcwdiluito1}  on stochastically stable
monomials.
\newline
Let us start the proof. Let us assume for a generic multi-overlap
monomial the following representation
$$
Q = \sum_{i_l^1}...\sum_{i_l^s}\prod_{l=1}^{n^a}\sigma_{i_l^a}^a
I(\{i_l^a \})
$$
where \textit{a} labels replicas, the inner product accounts for
the spins depicted by the index \textit{l} which belong to the
Boltzmann state \textit{a} of the product state $\Omega$ for the
multi-overlap $q_{a,a'}$ and runs over the integers from $1$ to
the amount of times the replica \textit{a} appears into the expression.
\newline
The external product multiplies all the terms coming from the
internal one. The factor $I$ fixes replica-bond constraints.
\newline
For example the monomial $Q = q_{12}q_{23}$ has $s=3, n^1=n^3=1,
n^2=2$ and $I=N^{-2}\delta_{i_1^1,i_1^2}\delta_{i_1^2,i_2^3}$,
there the $\delta$-functions give the correlations $1,2
\rightarrow q_{1,2}$ and $2,3 \rightarrow q_{2,3}$.
\newline
By applying the Boltzmann and quenched-disordered expectations we
get
$$
\langle Q \rangle_t = \textbf{E}\sum_{i_l^a}I(\{i_l^a
\})\prod_{a=1}^s \omega_{t}(\prod_{l=1}^{n^a}\sigma_{i_l^a}^a).
$$
Let us suppose now that  $Q$  is not stochastically stable (for
otherwise the proof will be simply ended) and let us decompose it
by factorizing the Boltzmann state $\omega$ and splitting  the
terms involving replicas appearing an even number of times from
the ones involving replicas appearing an odd number of times.
Then, evaluate the whole receipt at $t=1$.
$$
\langle Q \rangle_t = \textbf{E}\sum_{i_l^a,i_l^b}I(\{i_l^a \},
\{i_l^b \})\prod_{a=1}^u \omega_a (
\prod_{l=1}^{n^a}\sigma_{i_l^a}^a) \prod_{b=u+1}^s \omega_b (
\prod_{l=1}^{n^b}\sigma_{i_l^b}^b),
$$
where \textit{u} stands for the amount of replicas which appear an
odd number of times inside $Q$.
\newline
In this way we split the measure $\Omega$ in two ensembles
$\omega_a$ and $\omega_b$. Replicas belonging to $\omega_b$ are an
even number while the ones in $\omega_a$ an odd number.
\newline
At this point, as the Hamiltonian has two body interaction and
consequently is left unchanged by the symmetry  $\sigma_i^a
\rightarrow\sigma_i^a\sigma_{N+1}^a, \forall i \in (1,N)$ (as
$\sigma_{N+1}^2 \equiv 1$), we apply such a symmetry globally  to
the whole set of $N$ spins. The even measure is left unchanged by
this symmetry while the odd one takes a multiplying term
$\sigma_{N+1}$
$$
\langle Q \rangle = \sum_{i_l^a,i_l^b}I(\{i_l^a \}, \{i_l^b \})
\prod_{a=1}^u \omega ( \sigma_{N+1}^a
\prod_{l=1}^{n^a}\sigma_{i_l^a}^a) \prod_{b=u+1}^s \omega
(\sigma_{N+1}^b\prod_{l=1}^{n^b}\sigma_{i_l^b}^b).
$$
The last trick is that, by noticing the arbitrariness of the $N+1$
label in $\sigma_{N+1}$, we can change it to a generic label $k$
for each $k \neq \{ i_l^a \}$ and multiply by  $1=N^{-1}\sum_{k=1}^N$.
At finite $N$ the thesis is recovered forgetting terms $O(1/N)$
and becomes exact in the thermodynamic limit. $\Box$
\newline
\newline
It is straightforward to check that the effect of Theorem
\ref{Tcwdiluito1}  is not felt by stochastically stable
multi-overlap monomials (Corollary \ref{Tcwdiluito3}) thanks to
the dichotomy of the Ising spins ($\sigma_{N+1}^{2n}\equiv1
\forall n \in\mathbf{N} $). $\Box$
\newline
\newline
The last point missing to obtain the identities is finding a
streaming equation to work out the derivatives with respect to the
random field of the stochastically stable monomials. To this task
we introduce the following
\begin{proposition}\label{Tcwdfluw}
\textit{Given $F_s$ as a generic function of the spins of $s$
replicas, the following streaming equation holds}
\begin{eqnarray}\nonumber
\frac{\partial\langle F_s \rangle_{t,\tilde{\alpha}}}{\partial t}
&=& 2\tilde{\alpha}\theta [\sum_{a=1}^s\langle F_s \sigma_{i_0}^a
\rangle_{t,\tilde{\alpha}} - s \langle F_s \sigma_{i_0}^{s+1}
\rangle_{t,\tilde{\alpha}}] \quad +
\\ \nonumber
&+& \quad 2\tilde{\alpha}\theta^2 [ \sum_{a<b}^{1,s}\langle F_s
\sigma_{i_0}^a\sigma_{i_0}^b \rangle_{t,\tilde{\alpha}} - s
\sum_{a=1}^s\langle F_s \sigma_{i_0}^a\sigma_{i_0}^{s+1}
\rangle_{t,\tilde{\alpha}} + \\ \nonumber &+&
\frac{s(s+1)}{2!}\langle F_s \sigma_{i_0}^{s+1}\sigma_{i_0}^{s+2}
\rangle_{t,\tilde{\alpha}}] \quad +
\\ \nonumber
&+& \quad 2\tilde{\alpha}\theta^3 [\sum_{a<b<c}^{1,s}\langle F_s
\sigma_{i_0}^a\sigma_{i_0}^b\sigma_{i_0}^c
\rangle_{t,\tilde{\alpha}} - s \sum_{a<b}^{1,s}\langle F_s
\sigma_{i_0}^a\sigma_{i_0}^b\sigma_{i_0}^{s+1}
\rangle_{t,\tilde{\alpha}} +
\\ \nonumber
&+& \frac{s(s+1)}{2!}\sum_{a=1}^{s}\langle F_s \sigma_{i_0}^a
\sigma_{i_0}^{s+1}\sigma_{i_0}^{s+2} \rangle_{t,\tilde{\alpha}} +
\quad \\  \label{cwdfluw} &+& \frac{s(s+1)(s+2)}{3!} \langle F_s
\sigma_{i_0}^{s+1}\sigma_{i_0}^{s+2}\sigma_{i_0}^{s+3}
\rangle_{t,\tilde{\alpha}}] \, + \, O(\theta^3)
\end{eqnarray}
\end{proposition}
\medskip
\textbf{Proof}
\newline
The proof follows by direct calculations
\begin{eqnarray}
\frac{\partial\langle F_s \rangle_{t,\tilde{\alpha}}}{\partial t}
&=& \frac{\partial}{\partial t} \textbf{E} [
\frac{\sum_{\{\sigma\}}F_s
e^{\sum_{a=1}^s(\beta\sum_{\nu=1}^{P_{\tilde{\alpha} N}}
\sigma_{i_\nu}^a\sigma_{j_\nu}^a + \beta
\sum_{\nu=1}^{P_{2\tilde{\alpha }t}} \sigma_{i_\nu}^a)}}
{\sum_{\{\sigma\}}
e^{\sum_{a=1}^s(\beta\sum_{\nu=1}^{P_{\tilde{\alpha} N}}
\sigma_{i_\nu}^a\sigma_{j_\nu}^a + \beta
\sum_{\nu=1}^{P_{2\tilde{\alpha }t}} \sigma_{i_\nu}^a)}}] =
\\ \nonumber
&=& 2\tilde{\alpha} \textbf{E} [ \frac{\sum_{\{\sigma\}}F_s
e^{\sum_{a=1}^s(\beta\sigma_{i_0}^a +
\beta\sum_{\nu=1}^{P_{\tilde{\alpha} N}}
\sigma_{i_\nu}^a\sigma_{j_\nu}^a + \beta
\sum_{\nu=1}^{P_{2\tilde{\alpha }t}} \sigma_{i_\nu}^a)}}
{\sum_{\{\sigma\}} e^{\sum_{a=1}^s(\beta\sigma_{i_0}^a +
\beta\sum_{\nu=1}^{P_{\tilde{\alpha} N}}
\sigma_{i_\nu}^a\sigma_{j_\nu}^a + \beta
\sum_{\nu=1}^{P_{2\tilde{\alpha }t}} \sigma_{i_\nu}^a)}}] -
2\tilde{\alpha}\langle F_s \rangle_{t,\tilde{\alpha}} =
\\ \nonumber
&=& 2\tilde{\alpha} \textbf{E}[ \frac{\tilde{\Omega}_t (F_s
e^{\sum_{a=1}^s\beta\sigma_{i_0}^a})} {\tilde{\Omega}_t
(e^{\sum_{a=1}^s\beta\sigma_{i_0}^a})}] - 2\tilde{\alpha}\langle
F_s \rangle_{t,\tilde{\alpha}} =
\\ \nonumber
&=& 2\tilde{\alpha} \textbf{E}[ \frac{\tilde{\Omega}_t (F_s
\Pi_{a=1}^{s}(\cosh\beta + \sigma_{i_0}^a\sinh\beta))}
{\tilde{\Omega}_t (\Pi_{a=1}^{s}(\cosh\beta +
\sigma_{i_0}^a\sinh\beta))}] - 2\tilde{\alpha}\langle F_s
\rangle_{t,\tilde{\alpha}} =
\\ \nonumber
&=& 2\tilde{\alpha} \textbf{E}[ \frac{\tilde{\Omega}_t (F_s
\Pi_{a=1}^{s}(1 + \sigma_{i_0}^a\theta))} {(1 +
\tilde{\omega}_t(\sigma_{i_0}^a)\theta)^s}] -
2\tilde{\alpha}\langle F_s \rangle_{t,\tilde{\alpha}},
\end{eqnarray}
then, by noting that
$$
\Pi_{a=1}^{s}(1 + \sigma_{i_0}^a\theta) = 1 +
\sum_{a=1}^{s}\sigma_{i_0}^a\theta +
\sum_{a<b}^{1,s}\sigma_{i_0}^a\sigma_{i_0}^b\theta^2 +
\sum_{a<b<c}^{1,s}\sigma_{i_0}^a\sigma_{i_0}^b\sigma_{i_0}^c\theta^3
+ ...
$$
$$
\frac{1}{(1 + \tilde{\omega}_t \theta)^s} = 1 - s\tilde{\omega}_t
\theta + \frac{s(s+1)}{2!}\tilde{\omega}_t^2 \theta^2 -
\frac{s(s+1)(s+2)}{3!}\tilde{\omega}_t^3 \theta^3 + ...
$$
\medskip
\newline
we get
$$
\frac{\partial\langle F_s \rangle_{t,\tilde{\alpha}}}{\partial t}
= 2\tilde{\alpha} \textbf{E}[ \tilde{\Omega}_t (F_s(1 +
\sum_{a=1}^{s}\sigma_{i_0}^a\theta +
\sum_{a<b}^{1,s}\sigma_{i_0}^a\sigma_{i_0}^b\theta^2 +
\sum_{a<b<c}^{1,s}\sigma_{i_0}^a\sigma_{i_0}^b\sigma_{i_0}^c\theta^3
+ ...)) \times
$$
$$
\times (1 - s\tilde{\omega}_t \theta +
\frac{s(s+1)}{2!}\tilde{\omega}_t^2 \theta^2 -
\frac{s(s+1)(s+2)}{3!}\tilde{\omega}_t^3 \theta^3 + ...)] -
2\tilde{\alpha}\langle F_s \rangle_{t,\tilde{\alpha}},
$$
from which the thesis follows. $\Box$

\subsection{Linear constraints for multi-overlaps}

We saw that the stochastically stable multi-overlap monomials
become asymptotically independent by the $t$ parameter upon
increasing the size of the system. Calling for simplicity $G_N(q)$
a stochastically stable multi-overlap monomial, identities follow
as a consequence of Corollary \ref{Tcwdiluito3} and are encoded in
the following relation
$$
\lim_{N\to \infty} \partial_t \langle G_{N}(q) \rangle_t = 0.
$$
As we did when we investigated the Bernoullian model, we analyze
the stability of $\langle m^2 \rangle$ and $\langle
q_{12}^2\rangle$, up to the third order in $\theta$, so to compare
the results at the end.
\medskip
\newline
\begin{eqnarray}\nonumber
\partial_{t}\langle m_1^2 \rangle_t &=&
2\tilde{\alpha}\theta\Big (\langle m_1^3 \rangle_t - \langle
m_1^2m_2 \rangle_t\Big) - 2\tilde{\alpha}\theta^2 \Big(\langle
m_1^2q_{12} \rangle_t - \langle m_1^2q_{23} \rangle_t\Big) \\
\nonumber &+& \; 2\tilde{\alpha}\theta^3 \Big(\langle m_1^2q_{123}
\rangle_t - \langle m_1^2q_{234} \rangle_t\Big) + O(\theta^3)
\end{eqnarray}
\begin{eqnarray}\label{comp3}
\Rightarrow \quad && \Big[2\alpha\theta\Big (\langle m_1^4 \rangle
- \langle m_1^2m_2^2 \rangle\Big) - 2\alpha\theta^2
\Big( \langle m_1^2q_{12}^2 \rangle - \langle m_1^2q_{23}^2 \rangle\Big) \nonumber \\
&& + \; 2\alpha\theta^3 \Big(\langle m_1^2q_{123}^2 \rangle -
\langle m_1^2q_{234}^2 \rangle\Big) + O(\theta^3)\Big] = 0
\end{eqnarray}
\medskip
\begin{eqnarray}
\partial_{t}\langle q_{12}^2 \rangle_t &=& 4\tilde{\alpha}\theta
\Big(\langle m_1q_{12}^2 \rangle_t - \langle m_3q_{12}^2
\rangle_t\Big) + 2\tilde{\alpha}\theta^2 \Big(\langle q_{12}^3
\rangle_t - 4\langle
q_{12}^2q_{13}\rangle_t + 3\langle q_{12}^2q_{34} \rangle_t\Big) + \nonumber \\
&-& \quad 4\tilde{\alpha}\theta^3 \Big(\langle q_{12}^2q_{123}
\rangle_t - 3\langle q_{12}^2q_{134} \rangle_t + 4\langle
q_{12}^2q_{345} \rangle_t\Big) + O(\theta^3) \nonumber
\end{eqnarray}
\begin{eqnarray}\label{comp4}
\Rightarrow && \Big[4\alpha\theta \Big(\langle m_1^2q_{12}^2
\rangle - \langle m_3^2q_{12}^2 \rangle\Big) + 2\alpha\theta^2
\Big(\langle q_{12}^4 \rangle - 4\langle
q_{12}^2q_{13}^2\rangle + 3\langle q_{12}^2q_{34}^2 \rangle\Big) + \nonumber \\
&& - \quad 4\alpha\theta^3 \Big(\langle q_{12}^2q_{123}^2 \rangle
- 3\langle q_{12}^2q_{134}^2 \rangle + 2\langle q_{12}^2q_{345}^2
\rangle\Big) + O(\theta^3)\Big] = 0
\end{eqnarray}

\section{Discussion and outlook}

Let us start this section by comparing the results we get from the
two models. \newline We have to compare respectively eq.s
(\ref{aAC1}) versus
 (\ref{comp3}) and  (\ref{aAC2}) versus (\ref{comp4}).
 \newline
 We see that the details of the dilution do not affect the
 constraints: the series show the same set  of identities.
 \newline
In fact, despite we are
 not generally allowed to set to zero each term in the
 expressions (\ref{comp3},\ref{comp4},\ref{aAC1},\ref{aAC2}) (as we do to obtain alone the following identities (\ref{64}-\ref{69})),
at least close to the critical line, where different
multi-overlaps have different scaling laws \cite{ABC}, i.e.
$q_{n}^2 \propto (\alpha\theta - 1)^n$, such a spreading is
possible and we can forget each single
$(\alpha,\beta)$-coefficient as it does not affect the identities
(it is never involved into the averages $\langle . \rangle$).
\newline
We get
 \begin{eqnarray}\label{64}
0 &=& \langle m_1^4 \rangle - \langle m_1^2 m_2^2 \rangle, \\
\label{65} 0 &=& \langle m_1^2 q_{12}^2 \rangle - \langle
m_1^2q_{23}^2 \rangle, \\ \label{66} 0 &=& \langle m_1^2 q_{123}^2
\rangle - \langle m_1^2q_{234}^2 \rangle,
 \end{eqnarray}
 when investigating the magnetization as a trial function and
 \begin{eqnarray}\label{67}
0 &=& \langle m_1^2 q_{12}^2 \rangle - \langle m_3^2 q_{12}^2 \rangle, \\
\label{68} 0 &=& \langle q_{12}^4 \rangle - 4 \langle
q_{12}^2q_{23}^2 \rangle + 3 \langle q_{12}^2 q_{34}^2 \rangle, \\
\label{69} 0 &=& \langle q_{12}^2 q_{123}^2 \rangle - 3\langle
q_{12}^2 q_{134}^2 \rangle +2 \langle q_{12}^2 q_{345}^2 \rangle,
 \end{eqnarray}
 when investigating the two replica overlap.
\newline
Even if a minor point, we stress that the differences among the
global $(\alpha,\beta)$-coefficients (not shown here)
  clearly are related to the differences in the two method involved (in the former the constraints appear under the integral
over the temperature, while in the latter this $\beta$-average is
already worked out), furthermore the limiting connectivity in the
Bernoulli dilution is $\alpha/2$, while is $2\alpha$ in the
Poisson model; so there is an overall factor $4$ of difference
among the results (for the sake of clearness we worked out in the
appendix also the constraints in the Poisson diluted case via the
first method, to check explicitly the coherence among the two
model coefficients).
\newline
Then, by looking explicitly at the constraints, several physical
features can be recognized, in fact every term is well known:
\newline
the first class (Eq.s \ref{64},\ref{65},\ref{66}) is the standard
first momentum  self-averaging on replica symmetric systems; In
fact, by assuming replica equivalence, eq. (\ref{64}) turns out to
be the standard internal energy self-averaging of the Curie-Weiss
model. Eq.(\ref{65}) and (\ref{66}) contribute as higher order
internal energy self-average by taking into account the dilution
(in fact, they go to zero whenever $\alpha \to \infty$  because
$\theta$-powers higher than one go to zero and only the
Curie-Weiss self-averaging for the internal energy survives as it
should).
\newline
With a glance at the identities coming from the second constraint
series (Eq.s \ref{67},\ref{68},\ref{69}) we recognize immediately
the replica symmetry ansatz for the magnetization in the first
identity, followed by the first and the second Aizenman-Contucci
relation for systems with quenched disorder
\cite{aizcon}\cite{barra0}.
\newline
Interestingly these series are in agreement even with other
models, apparently quite far away, as  spin-glasses with Gaussian
coupling $\mathcal{N}[1,1]$ instead of $\mathcal{N}[0,1]$
\cite{CGN}. A very interesting conjecture may be that these
constraints hold for systems whose interaction has on average
positive strength and are affected by quenched disorder,
independently if the disorder affects the strength of the
interaction or the topology of the interaction. We plan to report soon on this topics.

\section*{Acknowledgment}

The author are pleased to thank an anonymous referee for useful
suggestions.
\newline
AB work in this paper is partially supported by the Technological
Vaucher Contract n.11606 of Calabria Region and partially by the
 CULTAPTATION Project (European Commission
contract FP6 - 2004-NEST-PATH-043434). PC acknowledge
Strategic Grant of University of Bologna.

\appendix

\section{Appendix}

\subsection{Alternative proof of energy self-averaging}

Starting from the thermodynamic relation
\begin{equation}\label{varh}
\textbf{E}[\Omega(h^2) - \Omega^2(h)] =
-\frac{1}{N}\frac{d}{d\beta}\textbf{E}[\Omega(h)]
\end{equation}
we evaluate explicitly the term $E[\Omega(h)]$ as
\begin{eqnarray}
\textbf{E}[\Omega(h)] &=& - \frac{1}{N}\textbf{E}\Big[
\frac{\sum_{\{\sigma\}}\sum_{\nu=1}^{k}
\sigma_{i_\nu}\sigma_{j_\nu}e^{-\beta H} }
{Z_N(\alpha,\beta)}\Big] \label{h1} =\\
&=& - \frac{1}{N}\textbf{E}\Big[ \frac{\sum_{\{\sigma\}}
k\sigma_{i_0}\sigma_{j_0}e^{-\beta H} }
{Z_N(\alpha,\beta)}\Big] \label{h2} =\\
&=& - \alpha'\textbf{E}\Big[ \frac{\sum_{\{\sigma\}}
\sigma_{i_0}\sigma_{j_0}e^{\beta\sigma_{i_0}\sigma_{j_0}}e^{-\beta
H} } {\sum_{\{\sigma\}}
e^{\beta\sigma_{k_0}\sigma_{l_0}}e^{-\beta H}}\Big] \label{h3} =\\
&=& - \alpha'\textbf{E}\Big[ \frac{\sum_{\{\sigma\}}
\sigma_{i_0}\sigma_{j_0}(\cosh\beta +
\sigma_{i_0}\sigma_{j_0}\sinh\beta)e^{-\beta H} }
{\sum_{\{\sigma\}}
(\cosh\beta + \sigma_{k_0}\sigma_{l_0}\sinh\beta)e^{-\beta H}}\Big] \label{h4} =\\
&=& - \alpha'\textbf{E}\Big[ \frac{\sum_{\{\sigma\}}
\sigma_{i_0}\sigma_{j_0}(1 +
\sigma_{i_0}\sigma_{j_0}\theta)e^{-\beta H} } {\sum_{\{\sigma\}}
(1 + \sigma_{k_0}\sigma_{l_0}\theta)e^{-\beta H}}\Big] \label{h5} =\\
&=& - \alpha'\textbf{E}\Big[
\frac{\omega(\sigma_{i_0}\sigma_{j_0}) + \theta} {1 +
\omega(\sigma_{k_0}\sigma_{l_0})\theta}\Big], \label{h6}
\end{eqnarray}
where in (\ref{h2}) we fixed the index  $\nu$, in (\ref{h3}) we
used the property (\ref{B1}) of the Bernoulli distribution and we
introduced two further families of random variables
$\{k_{\nu}\}$,$\{l_{\nu}\}$, and in (\ref{h4}) we used
$e^{\beta\sigma_{i_0}\sigma_{j_0}} = \cosh\beta +
\sigma_{i_0}\sigma_{j_0}\sinh\beta$.
\newline
Let us now expand the denominator of (\ref{h6}) keeping in mind
the relation
$$
\frac{1}{(1 + \tilde{\omega}_t \theta)^p} = 1 - p\tilde{\omega}_t
\theta + \frac{p(p+1)}{2!}\tilde{\omega}_t^2 \theta^2 -
\frac{p(p+1)(p+2)}{3!}\tilde{\omega}_t^3 \theta^3 + ...
$$
such that, by posing $p=1$, we obtain
\begin{equation}
\textbf{E}[\Omega(h)] = - \alpha'\textbf{E}\Big[ \theta +
\sum_{n=1}^{\infty}(-1)^n\theta^n(1-\theta^2)\langle
q_{1...n}^2\rangle \Big].
\end{equation}
By applying the modulus function to the equation above we can
proceed further with the following bound
\begin{equation}\label{breakdown}
|\textbf{E}[\Omega(h)]| \leq \alpha'\textbf{E}\Big[ |\theta| +
\sum_{n=1}^{\infty}|\theta^n(1-\theta^2)\langle
q_{1...n}^2\rangle| \Big].
\end{equation}
Both $|\theta|$ and $|\langle q_{1...n}^2\rangle|$  belong to
$[0,1]$ so we get
\begin{equation}
|E[\Omega(h)]| \leq \alpha'\Big[ 1 +
(1-\theta^2)\sum_{n=1}^{\infty}\theta^n \Big],
\end{equation}
whose harmonic series  converges to  $1/(1-\theta)$, $|\theta|<1$;
\newline
The fact that the convergence is not guaranteed at zero
temperature with this technique is not a problem because, first
the identities we are looking for hold in $\beta$-average,
secondly the zero temperature has been intensively investigated
elsewhere \cite{GDS}.
\newline
For each finite $\beta$, then,  we can write
\begin{eqnarray}
|\textbf{E}[\Omega(h)]| &\leq& \alpha'\Big[ 1 +
\frac{(1-\theta^2)}{1-\theta} \Big] = \\
&=& \alpha'\Big[ 1 +
\frac{(1-\theta)(1+\theta)}{1-\theta} \Big] = \\
&=&\alpha'\Big[ 1 +
(1+\theta) \Big] \leq \\
&\leq& 3\alpha', \label{minor}
\end{eqnarray}
and consequently
\begin{eqnarray}
\int_{\beta_1}^{\beta_2}\textbf{E}[\Omega(h^2) -
\Omega^2(h)]d\beta &\leq&
\int_{\beta_1}^{\beta_2}|\;\textbf{E}[\Omega(h^2) - \Omega^2(h)]\;|d\beta = \nonumber \\
&=& \frac{1}{N}\int_{\beta_1}^{\beta_2}|\;\frac{d}{d\beta}\textbf{E}[\Omega(h)]\;|d\beta \quad\leq \nonumber \\
&\leq& 3\frac{\alpha'}{N}
\end{eqnarray}
\begin{equation}
\Rightarrow \qquad \lim_{N\rightarrow\infty}
\int_{\beta_1}^{\beta_2}\textbf{E}[\Omega(h^2) -
\Omega^2(h)]d\beta = 0
\end{equation}
\medskip
and the proof is closed. $\Box$

\subsection{Details in Bernoulli dilution calculations}

Here some technical calculations concerning the self-averaging
technique on the Bernouilli diluted graph are reported.
\begin{eqnarray}
\textbf{E}[\Omega(h_lG)] &=& -\frac{1}{N}
\textbf{E}\Big[\frac{\sum_{\{\sigma\}}
\sum_{\nu=1}^{k}\sigma_{i_\nu}^l\sigma_{j_\nu}^lG
\:e^{(\beta\sum_{a=1}^{s}\sum_{\nu=1}^{k}\sigma_{i_\nu}^a\sigma_{j_\nu}^a)}}
{\sum_{\{\sigma\}}
e^{(\beta\sum_{a=1}^{s}\sum_{\nu=1}^{k}\sigma_{i_\nu}^a\sigma_{j_\nu}^a)}}] = \nonumber  \\
&=& -\frac{1}{N} \textbf{E}\Big[\frac{\sum_{\{\sigma\}} k
\sigma_{i_0}^l\sigma_{j_0}^lG
\:e^{(\beta\sum_{a=1}^{s}\sum_{\nu=1}^{k}\sigma_{i_\nu}^a\sigma_{j_\nu}^a)}}
{\sum_{\{\sigma\}}
e^{(\beta\sum_{a=1}^{s}\sum_{\nu=1}^{k}\sigma_{i_\nu}^a\sigma_{j_\nu}^a)}}],
\nonumber
\end{eqnarray}
and remembering the properties of the Bernoullian distribution
(\ref{B1}) we can write
\begin{eqnarray}
&=& -\frac{\alpha M}{N^2} \textbf{E}\Big[\frac{\sum_{\{\sigma\}}
\sigma_{i_0}^l\sigma_{j_0}^lG
\prod_{a=1}^{s}e^{\beta\sigma_{i_0}^a\sigma_{j_0}^a} \:e^{-\beta
H_s}} {\sum_{\{\sigma\}}
\prod_{a=1}^{s}e^{\beta\sigma_{i_0}^a\sigma_{j_0}^a} e^{-\beta
H_s}}\Big] = \nonumber  \\ \nonumber &=& -\frac{\alpha M}{N^2}
\textbf{E}\Big[\frac{\sum_{\{\sigma\}}
\sigma_{i_0}^l\sigma_{j_0}^lG \prod_{a=1}^{s}[\cosh\beta +
\sigma_{i_0}^a\sigma_{j_0}^a\sinh\beta] \:e^{-\beta H_s}}
{\sum_{\{\sigma\}} \prod_{a=1}^{s}[\cosh\beta +
\sigma_{i_0}^a\sigma_{j_0}^a\sinh\beta] \:e^{-\beta H_s}}\Big] =
\end{eqnarray}
\begin{eqnarray}
\nonumber  &=& -\frac{\alpha M}{N^2}
\textbf{E}\Big[\frac{\sum_{\{\sigma\}}
\sigma_{i_0}^l\sigma_{j_0}^lG \prod_{a=1}^{s}[1 +
\sigma_{i_0}^a\sigma_{j_0}^a\theta] \:e^{-\beta H_s}}
{\sum_{\{\sigma\}} \prod_{a=1}^{s}[1 +
\sigma_{i_0}^a\sigma_{j_0}^a\theta]
\:e^{-\beta H_s}}\Big] = \nonumber \\
&=& -\frac{\alpha M}{N^2}
\textbf{E}\Big[\frac{\Omega\Big(\sigma_{i_0}^l\sigma_{j_0}^lG
\prod_{a=1}^{s}[1 + \sigma_{i_0}^a\sigma_{j_0}^a\theta]\Big)}
{\Big(1 + \omega(\sigma_{i_0}^a\sigma_{j_0}^a)\theta\Big)^s}\Big]
= \nonumber
\end{eqnarray}
let us expand both the numerator and the denominator up to the
second order in $\theta$
\begin{eqnarray}
&&= -\frac{\alpha M}{N^2}
\textbf{E}\Big[\Omega\Big((\sigma_{i_0}^l\sigma_{j_0}^lG) (1 +
\sum_{a}^{1,s}\sigma_{i_0}^a\sigma_{j_0}^a\theta +
\sum_{a<b}^{1,s}\sigma_{i_0}^a\sigma_{j_0}^a\sigma_{i_0}^b\sigma_{j_0}^b\theta^2)\Big) \times \nonumber \\
&& \times \; \Big(1 - s\omega(\sigma_{i_0}\sigma_{j_0})\theta +
\frac{s(s+1)}{2}\omega^2(\sigma_{i_0}\sigma_{j_0})\theta^2\Big)\Big] = \nonumber \\
&&= -\frac{\alpha M}{N^2}
\textbf{E}\Big[\Omega\Big(G\sigma_{i_0}^l\sigma_{j_0}^l +
G\sum_{a}^{1,s}\sigma_{i_0}^a\sigma_{j_0}^a\sigma_{i_0}^l\sigma_{j_0}^l\theta
+
G\sum_{a<b}^{1,s}\sigma_{i_0}^a\sigma_{j_0}^a\sigma_{i_0}^b\sigma_{j_0}^b\sigma_{i_0}^l\sigma_{j_0}^l\theta^2\Big)
\times \nonumber
\end{eqnarray}
\begin{eqnarray}
&& \times \; \Big(1 - s\omega(\sigma_{i_0}^a\sigma_{j_0}^a)\theta +
\frac{s(s+1)}{2}\omega^2(\sigma_{i_0}\sigma_{j_0})\theta^2\Big)\Big] = \nonumber \\
&&= -\frac{\alpha M}{N^2}
\textbf{E}\Big[\Omega(G\sigma_{i_0}^l\sigma_{j_0}^l) +
\theta\Big(\sum_{a}^{1,s}\Omega(G\sigma_{i_0}^a\sigma_{j_0}^a\sigma_{i_0}^l\sigma_{j_0}^l)
-
s\Omega(G\sigma_{i_0}^l\sigma_{j_0}^l)\omega(\sigma_{i_0}\sigma_{j_0})\Big)
+ \nonumber \\
 && +
\;\theta^2\Big(\sum_{a<b}^{1,s}\Omega(G\sigma_{i_0}^a\sigma_{j_0}^a
\sigma_{i_0}^b\sigma_{j_0}^b\sigma_{i_0}^l\sigma_{j_0}^l) -
s\sum_{a}^{1,s}\Omega(G\sigma_{i_0}^a\sigma_{j_0}^a\sigma_{i_0}^l\sigma_{j_0}^l)
\omega(\sigma_{i_0}\sigma_{j_0}) + \nonumber
\end{eqnarray}
\begin{eqnarray}
&& \qquad + \quad \frac{s(s+1)}{2}\Omega(G\sigma_{i_0}^l\sigma_{j_0}^l)\omega^2(\sigma_{i_0}^a\sigma_{j_0}^a)\Big)\Big] = \nonumber \\
&&= -\frac{\alpha M}{N^2}\Big[ \langle Gm_l^2 \rangle +
\theta\Big(\sum_{a=1}^{s}\langle G q_{a,l}^2 \rangle -
s\langle G q_{l,s+1}^2 \rangle\Big) +  \nonumber \\
&&+  \theta^2\Big(\sum_{a<b}^{1,s}\langle Gq_{l,a,b}^2 \rangle -
s\sum_{a}^{1,s}\langle Gq_{l,a,s+1}^2 \rangle +
\frac{s(s+1)}{2}\langle Gq_{l,s+1,s+2}^2 \rangle\Big)\Big].
\nonumber
\end{eqnarray}
While the other term $\textbf{E}[\Omega(h_l)\Omega(G)]$ can be
worked out as follows: \be \textbf{E}[\Omega(h_l)\Omega(G)] = \ee
\begin{eqnarray}
&&=  -\frac{1}{N} \textbf{E}\Big[\frac{\sum_{\{\sigma\}}
\sum_{\nu=1}^{k}\sigma_{i_\nu}^l\sigma_{j_\nu}^lG
\:e^{\beta\sum_{\nu=1}^{k}\sigma_{i_\nu}^l\sigma_{j_\nu}^l}
\:e^{(\beta\sum_{a=1}^{s}\sum_{\nu=1}^{k}\sigma_{i_\nu}^a\sigma_{j_\nu}^a)}}
{\sum_{\{\sigma\}}
e^{\beta\sum_{\nu=1}^{k}\sigma_{i_\nu}^l\sigma_{j_\nu}^l}
e^{(\beta\sum_{a=1}^{s}\sum_{\nu=1}^{k}\sigma_{i_\nu}^a\sigma_{j_\nu}^a)}}] = \nonumber  \\
&&= -\frac{1}{N} \textbf{E}\Big[\frac{\sum_{\{\sigma\}} k
\sigma_{i_0}^l\sigma_{j_0}^lG
\:e^{\beta\sum_{\nu=1}^{k}\sigma_{i_\nu}^l\sigma_{j_\nu}^l}
\:e^{(\beta\sum_{a=1}^{s}\sum_{\nu=1}^{k}\sigma_{i_\nu}^a\sigma_{j_\nu}^a)}}
{\sum_{\{\sigma\}}
e^{\beta\sum_{\nu=1}^{k}\sigma_{i_\nu}^l\sigma_{j_\nu}^l}
e^{(\beta\sum_{a=1}^{s}\sum_{\nu=1}^{k}\sigma_{i_\nu}^a\sigma_{j_\nu}^a)}}]
= \nonumber
\end{eqnarray}
\begin{eqnarray}
&&= -\frac{\alpha M}{N^2} \textbf{E}\Big[\frac{\sum_{\{\sigma\}}
\sigma_{i_0}^l\sigma_{j_0}^lG
\:e^{\beta\sigma_{i_0}^l\sigma_{j_0}^l}
[\prod_{a=1}^{s}e^{\beta\sigma_{i_0}^a\sigma_{j_0}^a}] \:e^{-\beta
H_{s+1}}} {\sum_{\{\sigma\}} e^{\beta\sigma_{i_0}^l\sigma_{j_0}^l}
[\prod_{a=1}^{s}e^{\beta\sigma_{i_0}^a\sigma_{j_0}^a}]
e^{-\beta H_{s+1}}}\Big] = \nonumber  \\
&&= -\frac{\alpha M}{N^2}
\textbf{E}\Big[\frac{\Omega\Big(\sigma_{i_0}^l\sigma_{j_0}^lG (1 +
\sigma_{i_0}^l\sigma_{j_0}^l\theta) [\prod_{a=1}^{s}(1 +
\sigma_{i_0}^a\sigma_{j_0}^a\theta)]\Big)}
{\Big(1 + \omega(\sigma_{i_0}\sigma_{j_0})\theta\Big)^{s+1}}\Big] = \nonumber \\
&&= -\frac{\alpha M}{N^2}
\textbf{E}\Big[\Omega\Big((\sigma_{i_0}^l\sigma_{j_0}^lG +
G\theta) (1 + \sum_{a}^{1,s}\sigma_{i_0}^a\sigma_{j_0}^a\theta +
\sum_{a<b}^{1,s}\sigma_{i_0}^a\sigma_{j_0}^a\sigma_{i_0}^b\sigma_{j_0}^b\theta^2)\Big) \times \nonumber \\
&& \times \; \Big(1 - (s+1)\omega(\sigma_{i_0}\sigma_{j_0})\theta
+
\frac{(s+1)(s+2)}{2}\omega^2(\sigma_{i_0}\sigma_{j_0})\theta^2\Big)\Big]
= \nonumber
\end{eqnarray}
\begin{eqnarray}
&&= -\frac{\alpha M}{N^2}
\textbf{E}\Big[\Big(\Omega(G\sigma_{i_0}^l\sigma_{j_0}^l) +
\theta\Big(\Omega(G) + \sum_{a}^{1,s}\Omega(G\sigma_{i_0}^a\sigma_{j_0}^a\sigma_{i_0}^l\sigma_{j_0}^l)\Big) + \nonumber \\
&& + \:\theta^2\Big(\sum_{a}^{1,s}\Omega(G\sigma_{i_0}^a\sigma_{j_0}^a) + \sum_{a<b}^{1,s}\Omega(G\sigma_{i_0}^a\sigma_{j_0}^a\sigma_{i_0}^b\sigma_{j_0}^b\sigma_{i_0}^l\sigma_{j_0}^l)\Big) \times \nonumber \\
&& \times \; \Big(1 - (s+1)\omega(\sigma_{i_0}\sigma_{j_0})\theta
+
\frac{(s+1)(s+2)}{2}\omega^2(\sigma_{i_0}\sigma_{j_0})\theta^2\Big)\Big]
= \nonumber \end{eqnarray}
\begin{eqnarray}
&&= -\frac{\alpha M}{N^2}
\textbf{E}\Big[\Omega(G\sigma_{i_0}^l\sigma_{j_0}^l) +
\theta\Big(\Omega(G) +
\sum_{a}^{1,s}\Omega(G\sigma_{i_0}^a\sigma_{j_0}^a\sigma_{i_0}^l\sigma_{j_0}^l)
\nonumber
\\
&& -
(s+1)\Omega(G\sigma_{i_0}^l\sigma_{j_0}^l)\omega(\sigma_{i_0}\sigma_{j_0})\Big) + \nonumber \\
\end{eqnarray}
\begin{eqnarray} && +
\;\theta^2\Big(\sum_{a}^{1,s}\Omega(G\sigma_{i_0}^a\sigma_{j_0}^a)
+ \sum_{a<b}^{1,s}\Omega(G\sigma_{i_0}^a\sigma_{j_0}^a
\sigma_{i_0}^b\sigma_{j_0}^b\sigma_{i_0}^l\sigma_{j_0}^l)
\nonumber \\
&&-(s+1)\Omega(G)\omega(\sigma_{i_0}\sigma_{j_0}) + \nonumber \\
&& \quad - (s+1)\sum_{a}^{1,s}\Omega(G\sigma_{i_0}^a\sigma_{j_0}^a\sigma_{i_0}^l\sigma_{j_0}^l)
\omega(\sigma_{i_0}\sigma_{j_0})
\nonumber \\ && + \frac{(s+1)(s+2)}{2}\Omega(G\sigma_{i_0}^l\sigma_{j_0}^l)\omega^2(\sigma_{i_0}^a\sigma_{j_0}^a)\Big)\Big] = \nonumber
\end{eqnarray}
\begin{eqnarray}\nonumber
&&= -\frac{\alpha M}{N^2} \Big[\langle G m_{l}^2 \rangle +
\theta\Big(\langle G \rangle + \sum_{a=1}^{s}\langle G q_{a,l}^2
\rangle -
(s+1)\langle G q_{l,s+1}^2 \rangle\Big) + \\
&&\qquad + \quad \theta^2\Big(\sum_{a}^{1,s}\langle Gm_{a}^2 \rangle -
(s+1)\langle Gm_{l}^2 \rangle +
\sum_{a<b}^{1,s}\langle Gq_{l,a,b}^2 \rangle + \nonumber \\
&&\qquad - \quad (s+1)\sum_{a}^{1,s}\langle Gq_{l,a,s+1}^2 \rangle +
\frac{(s+1)(s+2)}{2}\langle Gq_{l,s+1,s+2}^2 \rangle\Big) \Big] \nonumber
\end{eqnarray}

\subsection{Poisson identities via the self-averaging technique}

For the sake of completeness we report also the constraints in the
Poisson diluted model obtained by using the first method:
\begin{eqnarray}\label{F1}
f_G(\alpha,\beta) &=& \alpha \Big[ \Big(\sum_{l=1}^{s}\langle G
m_l^2 \rangle - s\langle G m_{s+1}^2 \rangle\Big)
\Big(1 - \theta^2\Big) + \nonumber \\
&+& 2\theta \Big(\sum_{a<l}^{1,s}\langle G q_{al}^2 \rangle -
s\sum_{l}^{1,s}\langle G q_{l,s+1}^2 \rangle +
\frac{s(s+1)}{2}\langle G q_{s+1,s+2}^2 \rangle \Big) + \nonumber \\
&+& 3\theta^2\Big(\sum_{l<a<b}^{1,s}\langle G q_{l,a,b}^2 \rangle
- s\sum_{l<a}^{1,s}\langle G q_{l,a,s+1}^2 \rangle +
\frac{s(s+1)}{2}\sum_{l}^{1,s}\langle G q_{l,s+1,s+2}^2 \rangle + \nonumber \\
&-& \frac{s(s+1)(s+2)}{3!}\langle G q_{s+1,s+2,s+3}^2 \rangle
\Big)
 + O(\theta^3)\Big],
\end{eqnarray}
by which, choosing as a trial function $m^2$ we have
\begin{eqnarray}\nonumber
f_{m^2}^P(\alpha,\beta) &=& \alpha\Big[\Big(\langle m_1^4\rangle -
\langle m_1^2m_2^2\rangle\Big)\Big(1 - \theta^2\Big) + \\
\nonumber && - 2\theta\Big(\langle m_1^2q_{12}^2\rangle - \langle
m_1^2q_{23}^2\rangle\Big) + \\ \nonumber && + 3\theta^2
\Big(\langle m_1^2q_{123}^2\rangle - \langle m_1^2q_{234}^2\rangle
\Big) + O(\theta^3)\Big],
\end{eqnarray}
from which, changing the Jacobian $d\theta = (1 -
\theta^2)d\beta$, we get
\begin{eqnarray}\label{AC1}
\lim_{N\rightarrow\infty}\int_{\beta_1}^{\beta_2}
|f_{m^2}^P(\alpha,\beta)|d\beta &=&
\alpha\int_{\theta_1}^{\theta_2}d\theta \Big[|\;
\Big(\langle m_1^4\rangle - \langle m_1^2m_2^2\rangle\Big) + \\
\nonumber  &-& 2\frac{\theta}{(1 - \theta^2)} \Big(\langle
m_1^2q_{12}^2\rangle - \langle m_1^2q_{23}^2\rangle\Big) + \\
\nonumber &+& 3\frac{\theta^2}{(1 - \theta^2)} \Big(\langle
m_1^2q_{123}^2\rangle - \langle m_1^2q_{234}^2\rangle \\
\nonumber &+& O(\theta^3)\;|\Big] = 0.
\end{eqnarray}
\newline
\newline
If we set $G = q_{12}^2$ as the trial function
$f_G^P(\alpha,\beta)$ becomes
\begin{eqnarray} \nonumber
f_{q^2}^P(\alpha,\beta) &=& \alpha\Big[ \Big( 2\langle
m_1^2q_{12}^2\rangle - 2\langle m_3^2q_{12}^2\rangle\Big)\Big(1 -
\theta^2\Big) + \\ \nonumber &+& 2\theta\Big( \langle
q_{12}^4\rangle - 4\langle q_{12}^2q_{23}^2\rangle + 3\langle
q_{12}^2q_{34}^2\rangle\Big) + \\ \nonumber &-& 6\theta^2\Big(
\langle q_{12}^2q_{123}^2\rangle - 3\langle
q_{12}^2q_{234}^2\rangle + 2\langle q_{12}^2q_{345}^2\rangle\Big)
+ O(\theta^3)\Big].
\end{eqnarray}
Again
\begin{eqnarray}\label{AC2}
\lim_{N\rightarrow\infty}\int_{\beta_1}^{\beta_2}
|f_{q^2}^P(\alpha,\beta)|d\beta &=&
2\alpha\int_{\theta_1}^{\theta_2}d\theta \Big[|\; \Big(\langle
m_1^2q_{12}^2\rangle - \langle m_3^2q_{12}^2\rangle\Big) + \\
\nonumber &+& \frac{\theta}{(1 - \theta^2)}\Big( \langle
q_{12}^4\rangle - 4\langle q_{12}^2q_{23}^2\rangle + 3\langle
q_{12}^2q_{34}^2\rangle\Big) + \\ \nonumber &-&
3\frac{\theta^2}{(1 - \theta^2)}\Big( \langle
q_{12}^2q_{123}^2\rangle - 3\langle q_{12}^2q_{234}^2\rangle +
2\langle q_{12}^2q_{345}^2\rangle\Big) \nonumber \\ \nonumber &+&
O(\theta^3)\;|\Big] = 0,
\end{eqnarray}

\end{document}